# AN APPARENT GAP IN STELLAR MASS DISTRIBUTIONS AT $\approx 0.7\ M_\odot$ AND A POSSIBLE EXPLANATION

Robert L. Oldershaw

Physics Dept.

Amherst College

Amherst, MA 01002

USA

rlolders@unix.amherst.edu

## ABSTRACT

Mass functions for samples of white dwarf stars and for a large heterogeneous sample of nearby stars appear to have unexplained deficits in the $0.70 M_\odot$ to $0.75 M_\odot$ range. The existence, or non-existence, of this anomaly constitutes a definitive test of a fractal cosmological model that inherently predicts a gap in stellar mass functions at $\approx 0.73 M_\odot$.

## 1. INTRODUCTION

The goals of this paper are to introduce tentative empirical evidence for a possible anomaly in stellar mass distributions and to identify a potential explanation for the anomaly, in the event that it is verified by subsequent observations. It should be noted from the outset that most stellar mass determinations involve substantial uncertainties. Binary systems with stringent criteria can yield stellar masses that are accurate at the 2% level, but systems with these criteria are relatively rare. Andersen's (1991) high accuracy sample contains on the order of 100 stars, predominantly with masses greater than 1.0 $M_\odot$. Although stars with masses in the 0.1 $M_\odot$ to 1.0 $M_\odot$ range are present in huge numbers, their mass estimates usually have uncertainties of > 5%.

Typical stellar mass functions (SMFs) at the low-mass end of the mass spectrum tend to have a major peak in the 0.1 $M_\odot$ to 1.0 $M_\odot$ range, a fairly steep decline on the low-mass side of the peak, and a more gradual decline on the high-mass side. Given that expectation and maintaining an appreciation for stellar mass uncertainties, the author was impressed by an apparent gap at 0.70 $M_\odot$ – 0.75 $M_\odot$ in the mass function (MF) for a sample of white dwarf stars (Napiwotzki et al, 1999). Additional stellar mass function studies provided further support for the radical hypothesis that this deficit might be a universal property of stellar systems. The author's special interest in this hypothesis stems from his proposed cosmological model, the Self-Similar Cosmological Model (SSCM), which makes the definitive (i.e., unique, testable, prior and non-adjustable) prediction of a gap in the SMF at ≈ $0.73 M_\odot$ (Oldershaw, 1989a,b). If the global SMF does not have a significant deficit in the $0.70 M_\odot$ to $0.75 M_\odot$ range, then the SSCM will have been falsified. If, on the other hand, the predicted deficit is vindicated, then the SSCM will have passed a definitive test, and the nearly unavoidable conclusion would be that discrete cosmological self-similarity is one of nature's fundamental symmetries.

## 2. DATA

Figure 1 shows a histogram of mass values for a sample of 41 white dwarf stars analyzed by Napiwotzki et al. (1999) [NGS99]. The major peak between $0.50 M_\odot$ and $0.65 M_\odot$ is typical of white dwarf stars. Instead of a smooth decline at higher masses, however, the mass function drops precipitously beyond $0.65 M_\odot$, and the $0.70 M_\odot - 0.75 M_\odot$ bin is empty. There is a secondary peak at $0.75 M_\odot - 0.80 M_\odot$ with a reasonably smooth decline on the high-mass side. Although the small sample size limits what can be said with confidence about this mass function, it certainly *looks* like an anomalous deficit occurs between 0.70 $M_\odot$ and 0.75 $M_\odot$.

To test the reality of this deficit one would want a larger independent sample of white dwarfs for comparison. Fortunately NGS99 draw attention to a comparable data set by Bergeron et al. (1992) [BSL92], which contains three times as many stars. Figure 2 is a histogram of the BSL92 mass data for a non-overlapping sample of 127 white dwarf stars. Once again there is a predominant peak in the 0.50 $M_\odot$ to 0.65 $M_\odot$ range, and an apparent deficit at 0.70 $M_\odot$ to 0.75 $M_\odot$. As expected, there is a second peak at roughly 0.75 $M_\odot$ to 0.80 $M_\odot$ with a reasonably smooth tail at higher masses. Moreover, BSL92 show (their Fig. 13) that a comparable histogram of 68 white dwarf masses from the Palomar Green survey has a shape that is virtually identical to that of the BSL92 mass function in the 0.60 $M_\odot$ to 0.90 $M_\odot$ range.

When the NGS99 and BSL92 samples are combined (see Figure 3), one has a sample of 168 white dwarf stars. From a sharp peak at $\approx$ 0.55 $M_\odot$ the MF appears to fall off quite regularly with increasing mass, except for a gap between 0.65 $M_\odot$ and 0.75 $M_\odot$. It would require roughly 22 additional stars in these two bins in order to compensate for this deficit. It should be noted, however, that not all white dwarf mass distributions have similar deficits.

For example, there is at least one recent white dwarf sample in which a significant deficit at $\approx 0.7$ $M_\odot$ is not seen (Finley, *et al*, 1997). Ideally one would like a sample of $\approx 500$ white dwarf stars to check on the significance of this apparent anomaly. However such a sample with sufficiently accurate white dwarf masses is not presently available.

A natural question is whether the apparent deficit is seen only in white dwarf star samples, or whether it would also manifest itself in a heterogeneous sample of stars. Trying to find an appropriate sample of stars with which to test this question is not an easy matter, but there is an on-going NASA project whose aim is to accurately characterize the fundamental parameters of nearby stars (Blackman *et al*, 1998). At the website <http://nstars.arc.nasa.gov> one can find a histogram of mass values, plotted in 0.05 $M_\odot$ bins, for 319 of the stars nearest to Earth. Figure 4 reproduces the 0.05 $M_\odot$ – 1.0 $M_\odot$ portion (276 stars) of the NASA histogram. Since nearness to Earth is the major criterion for inclusion, the sample is heterogeneous. Although the deficit is somewhat less striking than the anomaly seen in Figure 3, and may be shifted to lower mass by 0.05 $M_\odot$, one sees a familiar shape in the mass distribution between 0.45 $M_\odot$ and 1.0 $M_\odot$. There is a peak at 0.50 $M_\odot$ – 0.55 $M_\odot$, a gradual decline to the 0.65 $M_\odot$ – 0.70 $M_\odot$ bin, and a rise to another peak in the 0.80 $M_\odot$ – 0.85 $M_\odot$ bin.

The statistical significance of the deficit at ≈ 0.70 $M_\odot$ for this sample may be low, but when viewed in the context of the white dwarf results, it raises an interesting possibility. It might be that MFs for many classes of stars have deficits in the 0.70 $M_\odot$ – 0.75 $M_\odot$ range because some physical mechanism inhibits star formation in this narrow mass range. Mass samples of sufficient size and accuracy to test this hypothesis are not currently available. Nevertheless the existing data are suggestive.

## 3. A REAL DEFICIT OR AN ARTIFACT?

Potential explanations for the observed deficit of stars at approximately 0.7 $M_\odot$ can be divided into two general categories: those that assume that the deficit is real and those that regard it as some sort of artifact. One simple explanation from the latter category is that the deficit is an artifact due to small sample size. The most striking example, in fact, is seen in the smallest sample (Napiwotzki et al, 1999). However, in all three samples the anomaly involves more than one bin. It also seems unlikely that virtually the same sample size artifact would be generated in two different samples of white dwarf stars totaling 168 stars and another heterogeneous sample of 276 stars. However, until larger samples are available, the small sample size issue cannot be fully dismissed.

Selection effects provide another possible class of explanations. But given the fact that the NSG99 and BSL92 samples have differing potential selection effects (Napiwotzki et al. 1999), the selection effect explanation is unlikely in the case of the white dwarf samples. In the case of the heterogeneous sample of 276 stars, there is a possible selection effect as noted by Green (2000). The class of K stars could be underrepresented because astronomers tend to find other classes of stars more interesting or more important to study. However, this explanation would not account for the deficits seen in the white dwarf samples.

A possible explanation for a real deficit at $\approx 0.7$ $M_\odot$ involves the merging of low-mass white dwarf stars to form high-mass white dwarfs (Napiwotzki, 2000). White dwarfs clearly have preferred masses in the 0.55 $M_\odot$ – 0.60 $M_\odot$ range, and merging would tend to produce a secondary peak at 1.10 $M_\odot$ – 1.20 $M_\odot$. Presumably there would be a detectable valley between the two peaks, although Napiwotzki (2000) has expressed doubts about whether $\approx$ 0.7 $M_\odot$ would be the likely location of the valley's center. Another candidate explanation for a real deficit is reserved for the next section.

## 4. A PREDICTED SMF GAP AT ≫ 0.73 $M_\odot$

The Self-Similar Cosmological Model (SSCM), which proposes that the cosmos has a discrete fractal organization, intrinsically predicts that there is a gap in the global SMF at ≈ 0.73 $M_\odot$. This fractal model has been formally presented in two review papers (Oldershaw 1989a, b). Here I will briefly outline the main ideas of the model and show how it unambiguously predicts a unique gap in stellar mass distributions.

If one takes a fresh look at the global properties of the cosmos, one is immediately struck by the highly stratified hierarchical organization of nature. The currently observable portion of the universe is comprised of galactic systems, which are comprised of stellar systems, which are comprised of atomic systems. The SSCM proposes that the atomic, stellar and galactic scales are three out of a large, and possibly infinite, number of nested cosmological scales.

If one studies systems from different scales, one sees evidence of discrete self-similarity wherein the n-scale "parts" have morphological, kinematical and dynamical properties that are analogous to those of n+1-scale systems. Two fairly overt examples are the self-similarity between atomic nuclei and neutron stars, and between micro-quasars and quasars. The author contends (Oldershaw, 1987, 1989a, b, 1996) that there is a remarkable incidence of unrecognized self-similarity among atomic, stellar and galactic scale systems, although this is masked by differences in spatial and temporal scales that exceed 17 orders of magnitude

Scale transformation equations that relate the length (R), time (T) and mass (M) measurements for self-similar analogues on neighboring scales n and n-1 are:

$$R_n = \mathbf{K} R_{n-1}, \qquad (1)$$

$$T_n = \mathbf{K} T_{n-1}, \qquad (2)$$

$$\text{and} \quad M_n = \mathbf{K}^{\mathbf{D}} M_{n-1}, \qquad (3)$$

where $\mathbf{K}$ ($\approx 5.2 \times 10^{17}$) and $\mathbf{D}$ ($\approx 3.174$) are dimensionless constants that have been determined empirically. These scale transformation equations have passed a battery of approximately 20 retrodictive tests (Oldershaw, 1989a). They have also led to definitive predictions, such as planets orbiting compact stellar objects, dominant dark matter populations at $\approx 0.2\ M_\odot$ and $\approx 0.6\ M_\odot$, and a steep decline in the global SMF below $0.15\ M_\odot$ (Oldershaw, 1996).

A defining principle of the SSCM is the general principle of discrete cosmological self-similarity, which asserts that for each type of system or fundamental property on a given cosmological scale, there is a self-similar analogue on all other scales. Quantitative measurements for systems and fundamental constants on differing scales are related by the scale transformation equations given above. The atomic scale has a highly distinctive "anomaly" in the distribution of atomic masses: there are no stable nuclei or isotopes with an atomic mass (A) of 5. Nuclei and isotopes can have A values of 1, 2, 3, 4, 6, 7..., but there is no A = 5. If the general principle of discrete cosmological self-similarity is valid, then the stellar scale must have an analogous gap, or at least a significant deficit, in the stellar mass function.

Given the mass of the proton and Eq. 3, one can calculate that the stellar equivalent to A = 1 has a value of $\approx 0.145$ $M_\odot$. Multiplying this value by 5 puts the stellar analogue to A = 5 at $\approx 0.725$ $M_\odot$. About a decade ago (Oldershaw 1989b) it was predicted that as stellar mass determinations become more accurate, similarities between atomic mass functions and stellar mass functions would become increasingly recognizable. Perhaps the deficit at 0.70 $M_\odot$ – 0.75 $M_\odot$, as discussed in section 2, is a tentative indication that this is the case. The typical SMF tends to have a primary peak at $\approx 0.15$ $M_\odot$ (Travis, 1994; Paresce, *et al*, 1995), and the white dwarf mass distribution of Napiwotzki et al (1999) has a very sharp peak at $\approx 0.589$ $M_\odot$; these peaks may correspond to the A = 1 and A = 4 peaks. Certainly it is premature to think that the present SMF evidence is sufficient to adequately test the hypothesis of self-similarity between stellar and atomic scale mass functions, but verification/falsification may be attainable within 5 to 10 years.

## 5. PREDICTIONS AND CONCLUSIONS

Tentative evidence for a deficit at $\approx 0.7$ $M_\odot$ has been found in several stellar mass functions. Larger samples and more accurate mass estimates for a variety of stellar classes will help to answer the following questions. Do white dwarf stars have a real MF deficit at $\approx 0.7$ $M_\odot$? Do other classes of stars have a similar deficit? If so, is there a full discontinuity or just a deficit at about 0.7 $M_\odot$?

Rarely does one have the situation where the presence or absence of a single testable phenomenon can verify or falsify a cosmological model. Yet this is the case for the proposed SMF deficit. The SSCM unambiguously predicts that *all* stellar mass distributions, if measured with uncertainties of < 5%, will manifest a significant deficit of stars with masses of $\approx 0.73$ $M_\odot$. No other theory known to the author predicts such a phenomenon, and successful retrodictions of the deficit by other models seem unlikely.

If the deficit is verified, then the SSCM will have passed a definitive test. In that case, whether there is a complete discontinuity at $\approx 0.73$ $M_\odot$, or merely an under-representation of stars in that mass range, will help in determining the degree of discrete cosmological self-similarity. A sharp discontinuity would argue for *exact* self-similarity, whereas a deficit would be more indicative of *statistical* self-similarity. If sufficiently large and accurate samples do not have deficits at $\approx 0.7$ $M_\odot$, then the SSCM will have been falsified.

## ACKNOWLEDGEMENTS

The author would like to thank Drs. Dana Blackman, Paul Green, Tod Henry, Ralf Napiwotzki, Rex Saffer, and Guillermo Torres for helpful comments on various aspects of stellar mass determinations.

| Range | Count |
|---|---|
| .05 - .1 | 0 |
| .1 - .15 | 0 |
| .15 - .2 | 0 |
| .2 - .25 | 0 |
| .25 - .3 | 0 |
| .3 - .35 | 0 |
| .35 - .4 | 0 |
| .4-.45 | 0 |
| .45-.5 | 1 |
| .5 - .55 | 4 |
| .55 - .6 | 13 |
| .6 - .65 | 12 |
| .65 - .70 | 1 |
| .70 -.75 | 0 |
| .75 - .8 | 5 |
| .8 - .85 | 2 |
| .85 - .9 | 2 |
| .9 - .95 | 0 |
| .95 - 1.0 | 1 |
| 1.0 - 1.05 | 0 |

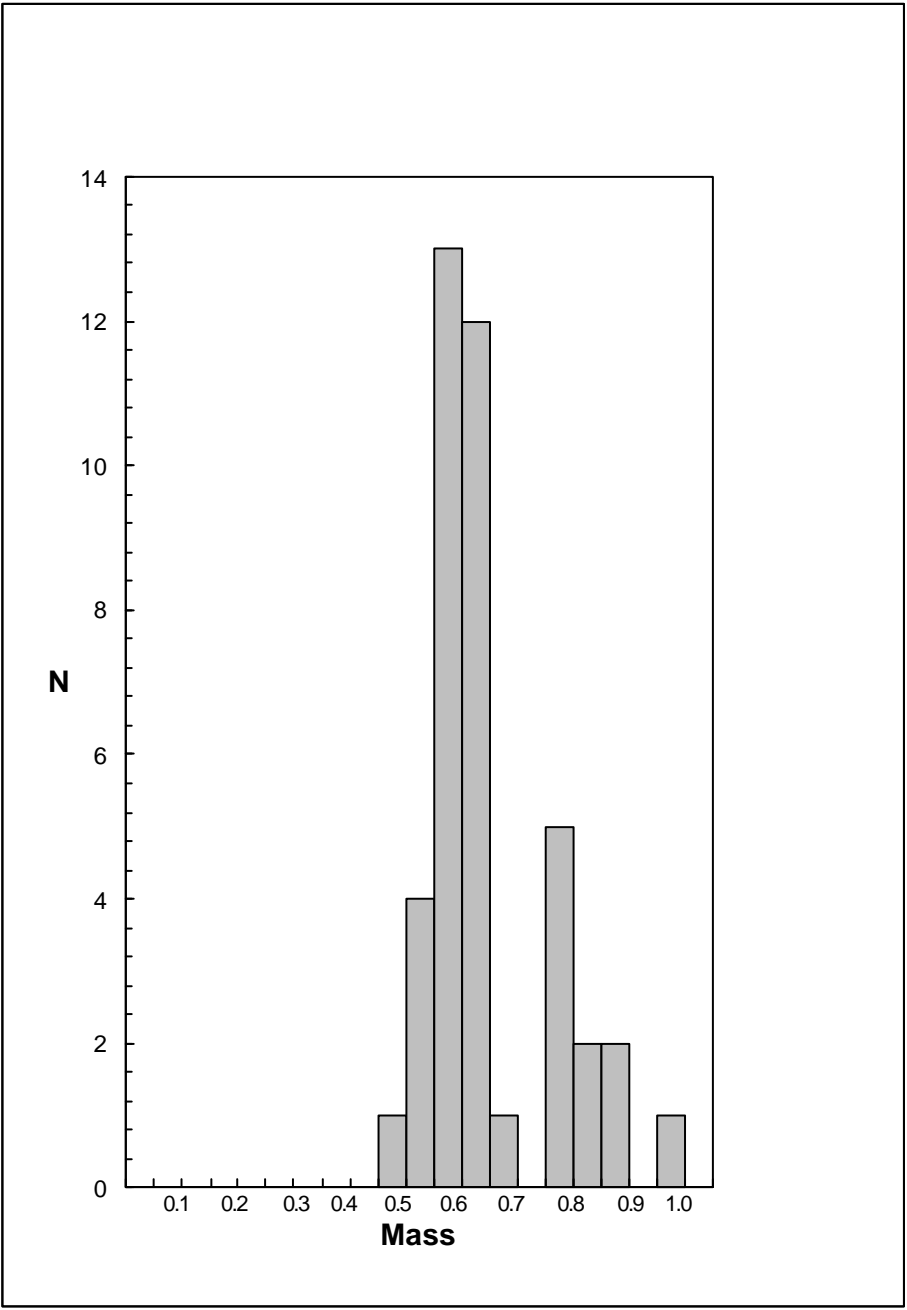

**Figure 1** Mass distribution for 41 white dwarf stars analyzed by Napiwotzki et al (1999).

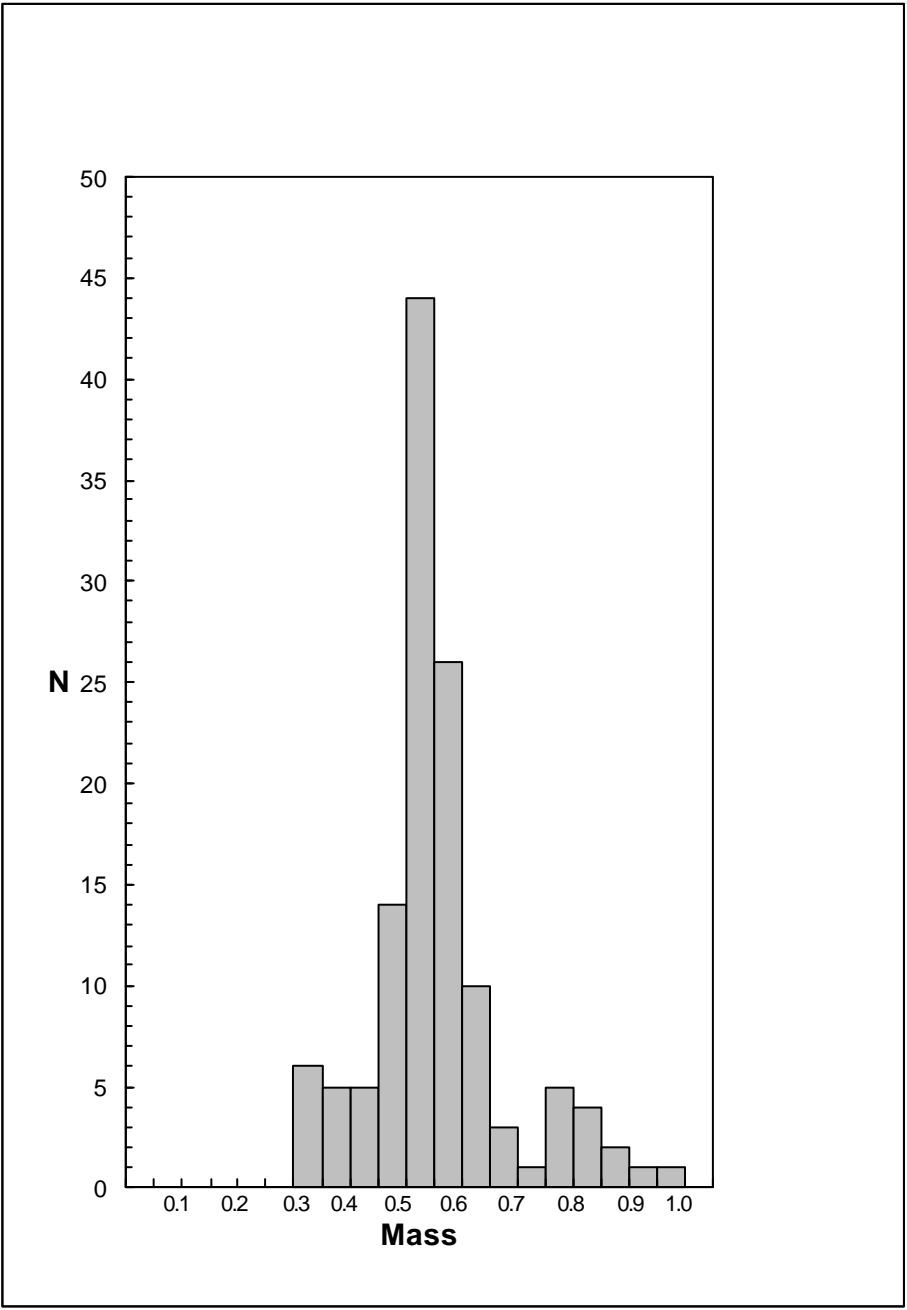

**Figure 2** Mass distribution for 127 white dwarf stars.
analyzed by Bergeron et al (1992).

| Range | N |
|---|---|
| .05 - .1 | 0 |
| .1 - .15 | 0 |
| .15 - .2 | 0 |
| .2 - .25 | 0 |
| .25 - .3 | 0 |
| .3 - .35 | 6 |
| .35 - .4 | 5 |
| .4-.45 | 5 |
| .45-.5 | 15 |
| .5 - .55 | 48 |
| .55 - .6 | 39 |
| .6 - .65 | 22 |
| .65 - .70 | 4 |
| .70 -.75 | 1 |
| .75 - .8 | 10 |
| .8 - .85 | 6 |
| .85 - .9 | 4 |
| .9 - .95 | 1 |
| .95 - 1.0 | 2 |
| 1.0 - 1.05 | 0 |

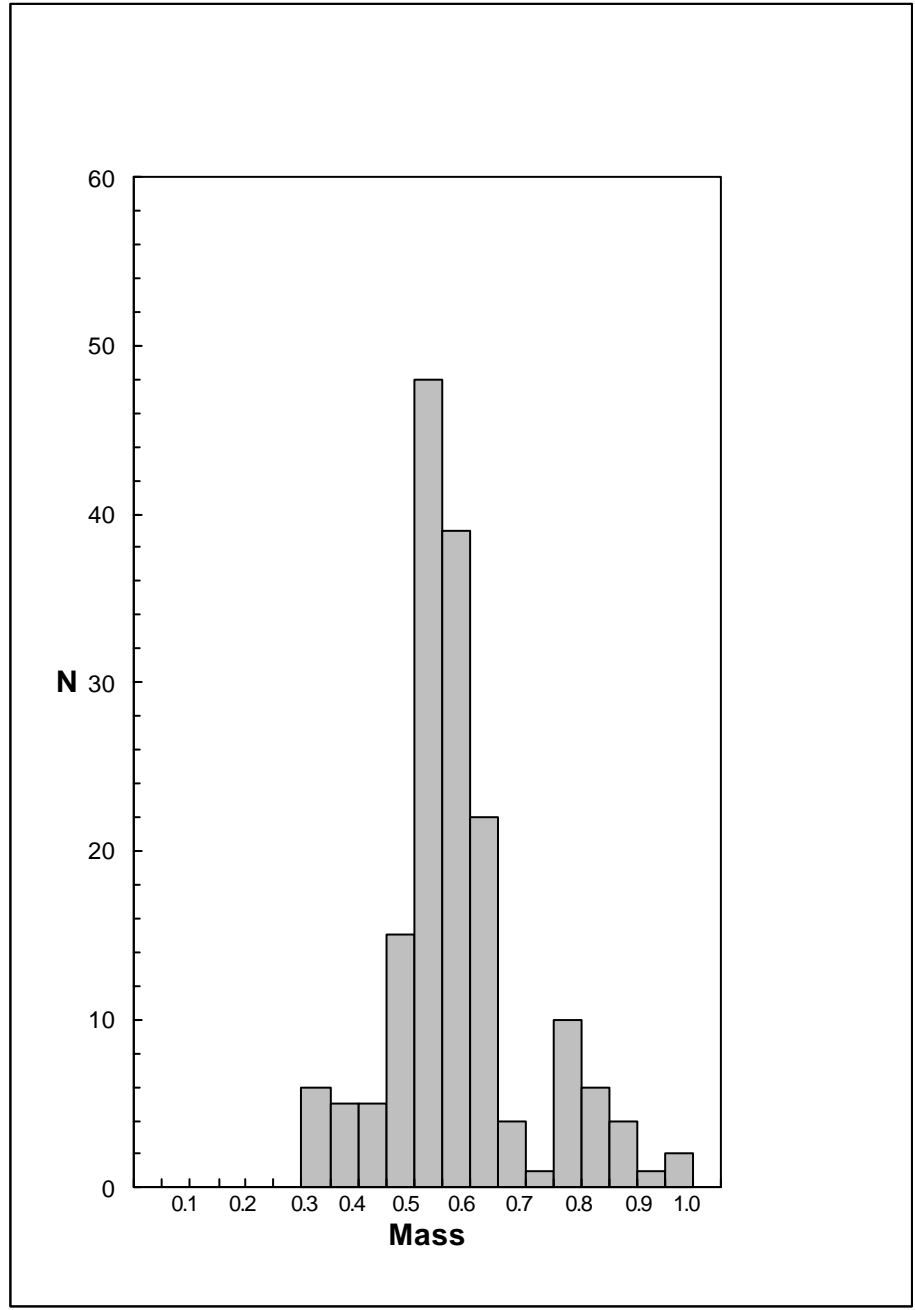

**Figure 3** Mass distribution for a sample of 168 white dwarf stars, formed by combining the samples of Napiwotzki et al (1999) and Bergeron et al (1992).

| Range | Count |
|---|---|
| .05 - .1 | 26 |
| .1 - .15 | 41 |
| .15 - .2 | 39 |
| .2 - .25 | 21 |
| .25 - .3 | 19 |
| .3 - .35 | 10 |
| .35 - .4 | 20 |
| .4 - .45 | 11 |
| .45 - .5 | 15 |
| .5 - .55 | 18 |
| .55 - .6 | 11 |
| .6 - .65 | 7 |
| .65 - .70 | 2 |
| .70 - .75 | 4 |
| .75 - .8 | 5 |
| .8 - .85 | 11 |
| .85 - .9 | 6 |
| .9 - .95 | 8 |
| .95 - 1.0 | 2 |
| 1.0 - 1.05 | 0 |

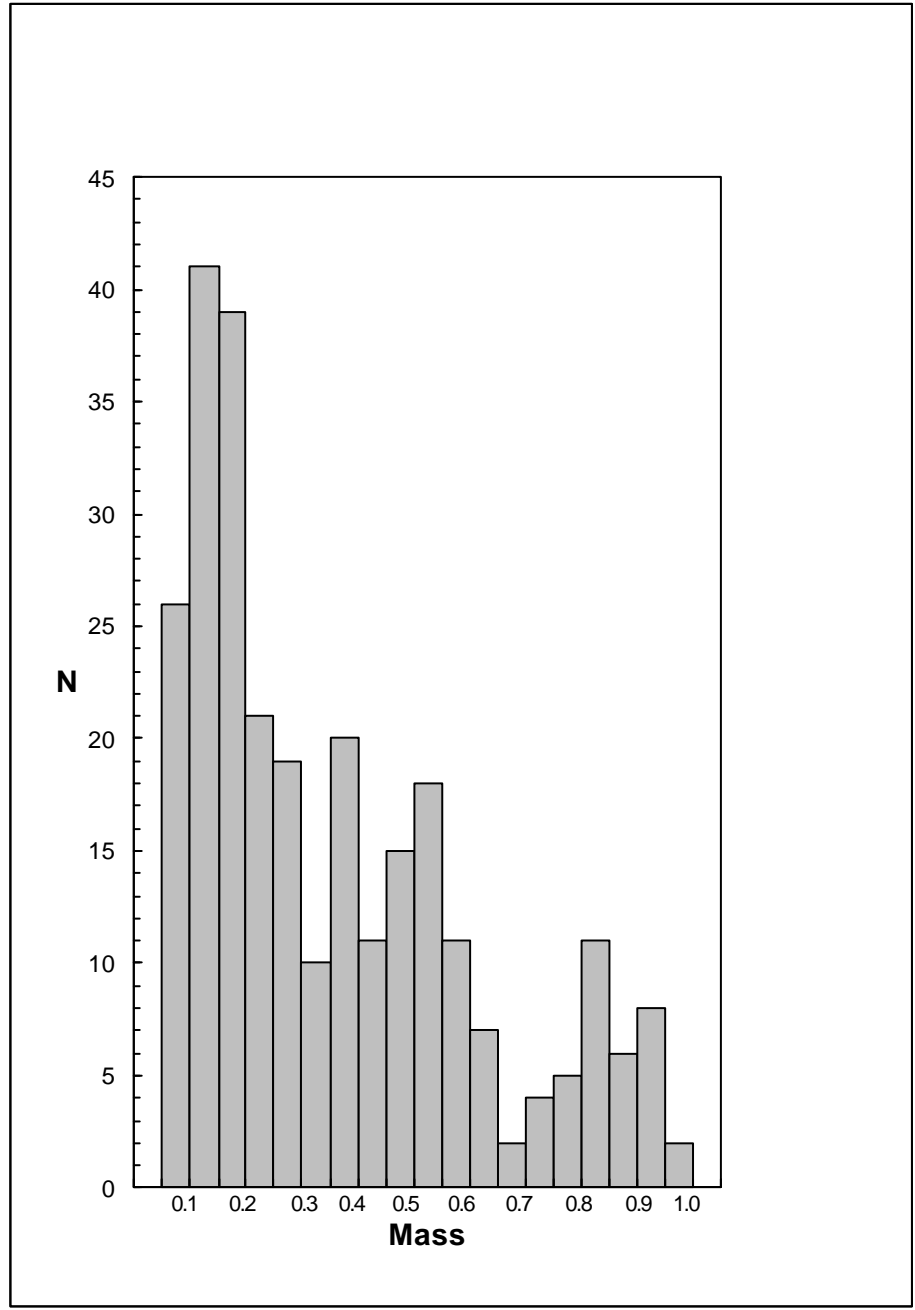

Figure 4  Mass distribution for a heterogeneous sample of 276 nearby stars in the range 0.05 $M_\odot$ to 1.0 $M_\odot$.  The data were collected by NASA's Nearby Stars Project team (see http://nstars.arc.nasa.gov).